\documentclass[sigconf,natbib=false]{acmart}

\usepackage{hyperref}
\usepackage{url}
\usepackage{graphicx}
\usepackage{multicol}
\usepackage{amsfonts}
\usepackage{amsmath}
\usepackage{amssymb}
\usepackage{booktabs}
\usepackage{amsthm}
\usepackage{multirow}
\usepackage{algpseudocode}
\usepackage{algorithm}
\usepackage{tabularx}

\usepackage{amsthm}

\theoremstyle{plain}

\theoremstyle{definition}

\AtBeginDocument{%
  }

\copyrightyear{2025}
\acmYear{2025}
\setcopyright{acmlicensed}
\acmConference[ICAIF '25]{6th ACM International Conference on AI in Finance}{November 15--18, 2025}{Singapore, Singapore}
\acmBooktitle{6th ACM International Conference on AI in Finance (ICAIF '25), November 15--18, 2025, Singapore, Singapore}
\acmDOI{10.1145/3768292.3770403}
\acmISBN{979-8-4007-2220-2/2025/11}

\RequirePackage[
  datamodel=acmdatamodel,
  style=acmnumeric,
  ]{biblatex}

\begin{document}

\title{Quantifying Semantic Shift in Financial NLP: Robust Metrics for Market Prediction Stability}


\author{Zhongtian Sun}
\email{zs256@kent.ac.uk}
\affiliation{%
  \institution{University of Kent \;|\; University of Cambridge}
  \city{}
  \country{United Kingdom}
}

\author{Chenghao Xiao}
\email{justinchenghaoxiao@gmail.com}
\affiliation{%
  \institution{Durham University}
  \city{}
  \country{United Kingdom}
}

\author{Anoushka Harit}
\email{ah2415@cam.ac.uk}
\affiliation{%
  \institution{University of Cambridge}
  \city{}
  \country{United Kingdom}
}

\author{Jongmin Yu}
\email{jmandrewyu@gmail.com}
\affiliation{%
  \institution{ProjectG.AI}
  \city{}
  \country{Republic of Korea}
}

\renewcommand{\shortauthors}{Sun et al.}

\begin{abstract}
Financial news is essential for accurate market prediction, but evolving narratives across macroeconomic regimes introduce semantic and causal drift that weaken model reliability. We present an evaluation framework to quantify robustness in financial NLP under regime shifts. The framework defines four metrics: (1) Financial Causal Attribution Score (FCAS) for alignment with causal cues, (2) Patent Cliff Sensitivity (PCS) for sensitivity to semantic perturbations, (3) Temporal Semantic Volatility (TSV) for drift in latent text representations, and (4) NLI-based Logical Consistency Score (NLICS) for entailment coherence. Applied to LSTM and Transformer models across four economic periods (pre-COVID, COVID, post-COVID, and rate hike), the metrics reveal performance degradation during crises. Semantic volatility and Jensen-Shannon divergence correlate with prediction error. Transformers are more affected by drift, while feature-enhanced variants improve generalisation. A GPT-4 case study confirms that alignment-aware models better preserve causal and logical consistency. The framework supports auditability, stress testing, and adaptive retraining in financial AI systems.
\end{abstract}

\begin{CCSXML}
<ccs2012>
   <concept>
       <concept_id>10010147.10010178.10010187.10010190</concept_id>
       <concept_desc>Computing methodologies~Probabilistic reasoning</concept_desc>
       <concept_significance>500</concept_significance>
       </concept>
   <concept>
       <concept_id>10010147.10010178.10010187.10010198</concept_id>
       <concept_desc>Computing methodologies~Reasoning about belief and knowledge</concept_desc>
       <concept_significance>300</concept_significance>
       </concept>
   <concept>
       <concept_id>10010147.10010257.10010293.10010300</concept_id>
       <concept_desc>Computing methodologies~Learning in probabilistic graphical models</concept_desc>
       <concept_significance>500</concept_significance>
       </concept>
 </ccs2012>
\end{CCSXML}

\ccsdesc[500]{Computing methodologies~Natural language processing}
\ccsdesc[500]{Computing methodologies~Machine learning}
\ccsdesc[500]{Computing methodologies~Neural networks}
\ccsdesc[500]{Computing methodologies~Causal inference}

\keywords{Financial Natural Language Processing, Semantic Drift, Causal Inference, Regime Shift Robustness}

\maketitle

\section{Introduction}
Financial markets are acutely sensitive to news, and predictive models increasingly rely on textual signals to forecast asset returns. However, financial narratives evolve significantly over time due to macroeconomic shocks, policy shifts, and global crises. These changes introduce \textit{semantic and causal shifts} in how events are framed and how they relate to asset price movements. For instance, the onset of COVID19 fundamentally altered both the tone and the causal structure of financial news, often misaligning model assumptions with real-world outcomes.

While previous work in financial NLP has addressed sentiment classification~\cite{du2024financial}, event extraction~\cite{ding2015deep}, and domain-specific language modelling~\cite{sinha2022sentfin}, limited attention has been given to how predictive performance deteriorates under distributional and semantic drift. In this study, we focus on \textbf{stock return prediction} and introduce a structured evaluation framework that quantifies model robustness across distinct economic regimes.

We propose four complementary diagnostic metrics:

\begin{itemize}
    \item \textbf{FCAS} (Financial Causal Attribution Score): Measures alignment between model predictions and implied causal statements in financial news.
    \item \textbf{PCS} (Patent Cliff Sensitivity): Assesses the effect of controlled semantic perturbations on prediction stability.
    \item \textbf{TSV} (Temporal Semantic Volatility): Quantifies shifts in latent text representations over time.
    \item \textbf{NLICS} (NLI-based Logical Consistency Score): Evaluates the coherence between predictions and text using natural language inference.
\end{itemize}
We evaluate LSTM and Transformer models across four macroeconomic regimes: \textit{pre COVID}, \textit{COVID}, \textit{post COVID}, and the \textit{rate hike period}, capturing narrative and temporal variation~\cite{wu2020outbreak}. Our findings show that semantic drift, measured via Jensen-Shannon divergence (with a peak of \textbf{0.24} between COVID and the rate hike period), is strongly associated with elevated prediction error. LSTM models exhibit greater stability across regimes, while Transformers display heightened sensitivity to narrative shifts. Feature-enhanced Transformers reduce semantic volatility and improve generalisation.\\

\noindent\textbf{Our key contributions are as follows:}
\begin{enumerate}
    \item We introduce a novel framework for evaluating the robustness of financial NLP models under macroeconomic regime shifts, grounded in causal and semantic diagnostics.
    \item We propose four metrics (FCAS, PCS, TSV, NLICS) to assess causal alignment, perturbation sensitivity, semantic volatility, and logical coherence of model predictions.
    \item We conduct a comprehensive empirical analysis on LSTM and Transformer models across four economic phases, revealing how different architectures respond to semantic drift.
    \item We provide a GPT4-based case study to validate our metrics using entailment-based reasoning, offering insights into model alignment and trustworthiness.
    \item Our framework enables model auditability, stress testing, and adaptive retraining, addressing the needs of robust AI deployment in real-world financial settings.
\end{enumerate}

\section{Related Work}
% \textbf{Financial NLP.} Prior studies have applied sentiment-based models to forecast stock movements~\cite{wu2023bloomberggpt}, yet these often overlook how temporal shifts in language affect model robustness. Transformer-based architectures, such as FinBERT~\cite{araci2019finbert}, have demonstrated improved performance by incorporating domain-specific financial corpora. However, they still face limitations under evolving macroeconomic conditions.

% \textbf{Semantic Shift in NLP.} Work in domain adaptation has investigated language drift across contexts~\cite{gururangan2020don}, but financial discourse poses distinct challenges due to abrupt and structural regime changes. While studies on linguistic volatility and representation drift~\cite{chang2023characterizing} offer valuable tools, their application to financial forecasting remains limited.

% \textbf{Logical Consistency in LLMs.} Recent advances in natural language inference have enabled large language models to assess logical coherence in generated outputs~\cite{ghosh2024logical}. Despite progress in this area, the use of NLI-based evaluation in financial prediction, particularly under causal and semantic drift, remains largely unexplored.

\textbf{Financial NLP.}  
Prior studies have applied sentiment-based models to forecast stock movements~\cite{wu2023bloomberggpt}, yet these often overlook how temporal shifts in language affect model robustness. Transformer-based architectures, such as FinBERT~\cite{araci2019finbert, harit2024breaking}, have demonstrated improved performance by incorporating domain-specific financial corpora. FinBERT continues to be a foundation in financial sentiment analysis, with multiple studies validating its superior performance over general-purpose models~\cite{priya2025advanced,mahendran2025comparative}. Recent work also shows that GPT-4o, when prompt-optimized, can outperform FinBERT in financial news sentiment tasks, especially across sectors~\cite{sun2023money,kang2025comparative, sun2025ricciflowrec}. Hybrid approaches, such as FinBERT-LSTM and GPT-augmented TinyFinBERT, further improve prediction accuracy while addressing model efficiency~\cite{jun2024predicting,thomas2024enhancing}. However, despite architectural advances, these models often struggle to generalize under distributional shifts introduced by volatile macroeconomic conditions.

\textbf{Semantic Shift in NLP.}  
Work in domain adaptation has investigated language drift across contexts~\cite{gururangan2020don, sun2022contrastive}, but financial discourse poses distinct challenges due to abrupt and structural regime changes. While studies on linguistic volatility and representation drift offer valuable tools, their application to financial forecasting remains limited. Recent work demonstrates that public sentiment volatility, measured via FinBERT embeddings, can strongly affect systemic risk forecasting~\cite{jin2025early}. Domain-specific pretraining remains a debated strategy: continual pretraining from general models often outperforms full in-domain pretraining under real-world drift scenarios~\cite{peng2021domain}. Moreover, large-scale evaluations confirm transformer models like FinBERT are more robust than traditional methods for classifying large, heterogeneous financial corpora~\cite{correa2022neural}.

\textbf{Logical Consistency in LLMs.}  
Recent advances in natural language inference have enabled large language models to assess logical coherence in generated outputs~\cite{ghosh2024logical}. However, most evaluation studies focus on generic benchmarks and overlook domain-specific logical robustness~\cite{harit2025textfold}. In financial contexts, logical consistency is critical under volatile narratives, yet few benchmarks address this. Studies have shown that BERT-based models can outperform GPT variants on financial reasoning tasks due to better interpretability and reduced hallucination risk~\cite{sharkey2024bert}. Similar trade-offs have been observed in direct comparisons: while GPT-4 offers higher flexibility, FinBERT often yields more stable and interpretable outputs in sentiment-based market prediction tasks~\cite{shobayo2024innovative}. Fine-tuned variants of FinBERT, designed to resolve sentiment ambiguity in complex sentences, also demonstrate improved logical alignment~\cite{gossi2023finbert,sun2025glance}. Despite this, structured NLI-based evaluation for drift-aware logical robustness in finance remains underexplored.

\section{Problem Formulation}
Let $\mathcal{X}$ denote the space of financial text inputs (e.g., news articles), and $\mathcal{Y} \subset \mathbb{R}$ the space of stock return targets. A model $f_\theta: \mathcal{X} \to \mathcal{Y}$ with parameters $\theta$ aims to minimise prediction error on input–output pairs $(x, y) \sim \mathcal{P}$, where $\mathcal{P}$ is the underlying joint distribution.

In practice, financial data are non-stationary, with distribution $\mathcal{P}$ changing across macroeconomic regimes. Let $\mathcal{R} = \{r_1, \ldots, r_K\}$ denote a finite set of regimes (e.g., pre-COVID, COVID, post-COVID, rate-hike). Each regime $r_k$ induces a distinct distribution $\mathcal{P}_{r_k}$ over $\mathcal{X} \times \mathcal{Y}$.

\paragraph{Objective.}
We aim to learn a predictor $f_\theta$ that achieves not only low expected error under each $\mathcal{P}_{r_k}$, but also exhibits robustness to distributional shift and semantic drift across regimes. The standard learning objective minimises the expected squared loss:
\begin{equation}
\mathcal{L}_{\text{MSE}}(\theta; \mathcal{P}_{r_k}) = \mathbb{E}_{(x,y) \sim \mathcal{P}_{r_k}} \left[ \left( f_\theta(x) - y \right)^2 \right]
\end{equation}

However, $\mathcal{L}_{\text{MSE}}$ alone fails to capture structural misalignments introduced by temporal or causal shift in financial language. To evaluate robustness, we define four metric functionals, each mapping a model–data pair $(f_\theta, \mathcal{P}_{r_k})$ to a real value:

\begin{align}
\text{FCAS}(f_\theta, \mathcal{P}_{r_k}) &= \mathbb{E}_{(x, c)} \left[ \mathbb{I} \left( \operatorname{sign}(f_\theta(x)) = \operatorname{sign}(c) \right) \right] \\
\text{PCS}(f_\theta, \mathcal{P}_{r_k}) &= \mathbb{E}_{x \sim \mathcal{P}_{r_k}} \left[ f_\theta(x) - f_\theta(\tilde{x}) \right] \\
\text{TSV}(f_\theta, \mathcal{P}_{r_k}) &= \mathbb{E}_{(x_t, x_{t+1})} \left[ \left\| \phi(x_{t+1}) - \phi(x_t) \right\|_2 \right] \\
\text{NLICS}(f_\theta, \mathcal{P}_{r_k}) &= \mathbb{E}_{x \sim \mathcal{P}_{r_k}} \left[ \operatorname{EntailmentScore}(x, \mathcal{H}(f_\theta(x))) \right]
\end{align}

Here, $c$ denotes an extracted causal claim, $\tilde{x}$ is a counterfactual perturbation of $x$, and $\phi(x)$ denotes a sentence embedding. $\mathcal{H}(f_\theta(x))$ is a natural language hypothesis constructed from the model prediction (e.g., “The stock will rise”). The entailment score is computed using a pretrained natural language inference model.

\paragraph{Robustness Profile.}
We define the regime-wise robustness profile as a vector:
\begin{equation}
\mathcal{M}_{r_k}(f_\theta) = \left[ \mathcal{L}_{\text{MSE}},\ \text{FCAS},\ \text{PCS},\ \text{TSV},\ \text{NLICS} \right](f_\theta, \mathcal{P}_{r_k})
\end{equation}

\textbf{Goal.}
Our aim is not to minimise these metrics jointly, but to provide a diagnostic framework. By analysing $\{ \mathcal{M}_{r_k} \}_{k=1}^K$, we characterise failure modes of $f_\theta$ under temporal and narrative shift, and identify when model retraining, adaptation, or rejection may be warranted.

\section{Methodology}
We introduce a diagnostic framework for evaluating financial NLP models under distributional and semantic shift. The framework integrates predictive modelling with four complementary metrics designed to capture causal alignment, perturbation robustness, semantic volatility, and logical coherence.

\begin{figure}[ht]
    \centering
    \includegraphics[width=\linewidth]{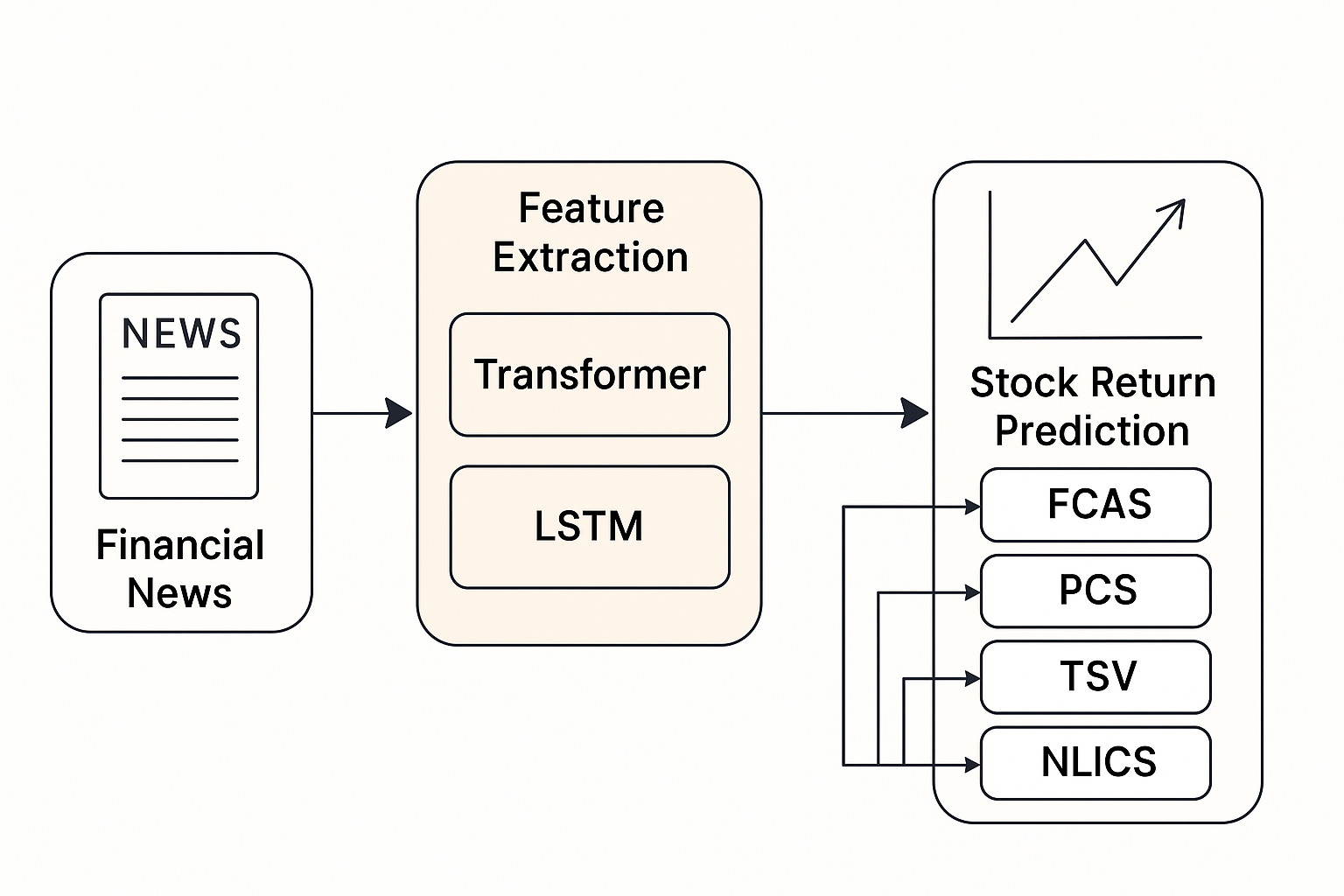}  
    \caption{Regime-aware evaluation framework. Financial news is encoded via an LSTM or Transformer, which produces stock return predictions. Predictions are evaluated using four diagnostic metrics across macroeconomic regimes.}
    \Description{Diagram of the evaluation pipeline: news text enters an encoder (either LSTM or Transformer), producing embeddings; a prediction head outputs stock return forecasts; the outputs are assessed by four metrics (FCAS, PCS, TSV, NLICS) across different macroeconomic regimes.}
    \label{fig:architecture}
\end{figure}

Figure~\ref{fig:architecture} summarises the workflow. Given a financial news article, features are extracted using either an LSTM or Transformer-based encoder. The model output (predicted next-day stock return) is evaluated with four diagnostic metrics: Financial Causal Attribution Score (FCAS), Patent Cliff Sensitivity (PCS), Temporal Semantic Volatility (TSV), and NLI-based Logical Consistency Score (NLICS). Evaluating these metrics across distinct macroeconomic regimes enables a systematic characterisation of robustness under structural and narrative change.

\subsection{Task Definition}
Let $\mathcal{X}$ denote the space of financial news inputs and $\mathcal{Y} \subset \mathbb{R}$ the space of next-day stock returns. Each datapoint $(x_i, y_i)$ pairs an article $x_i$ with its realised return $y_i$. The prediction function $f_\theta: \mathcal{X} \rightarrow \mathcal{Y}$ is trained to minimise mean squared error:
\begin{equation}
\mathcal{L}_{\text{MSE}}(\theta) = \frac{1}{N} \sum_{i=1}^N \big(f_\theta(x_i) - y_i\big)^2.
\end{equation}

Because financial narratives evolve over time, we partition the dataset into $K$ macroeconomic regimes $\mathcal{R} = \{r_1, \ldots, r_K\}$, each inducing a distinct distribution $\mathcal{P}_{r_k}$ over $(x, y)$.

\subsection{Model Architectures}
We evaluate three representative model classes:
\begin{itemize}
    \item \textbf{LSTM}: trained on TF–IDF vectors to capture sequential dependencies in bag-of-words style inputs.
    \item \textbf{Transformer}: a fine-tuned DistilBERT encoder applied directly to raw financial text.
    \item \textbf{Feature-based Transformer}: combines TF–IDF vectors with MiniLM sentence embeddings, providing both sparse and dense representations.
\end{itemize}
All models are implemented in PyTorch and trained using Adam (learning rate 0.001, batch size 64, hidden size 256, dropout 0.2).

\subsection{Evaluation under Regime Shift}
For each regime $r_k$, robustness is quantified via a vector of functional metrics $\mathcal{M}_{r_k}(f_\theta)$. These go beyond accuracy by probing causal alignment, semantic fragility, temporal drift, and logical consistency:

\paragraph{Financial Causal Attribution Score (FCAS).}
Measures whether the predicted return direction agrees with causal claims extracted from the article:
\begin{equation}
\text{FCAS} = \frac{1}{N} \sum_{i=1}^{N} \mathbb{I}\!\left[\operatorname{sign}(f_\theta(x_i)) = \operatorname{sign}(c_i)\right],
\end{equation}
where $c_i$ denotes the polarity of extracted causal cues. This captures whether models align with narrative drivers reported in the text.

\paragraph{Perturbation Sensitivity (PCS).}
Assesses robustness to small but meaningful linguistic changes, analogous to stress-testing in risk management:
\begin{equation}
\text{PCS} = \mathbb{E}_{x_i} \Big[f_\theta(x_i) - f_\theta(\tilde{x}_i)\Big],
\end{equation}
where $\tilde{x}_i$ is a perturbed version of $x_i$ (e.g., replacing “growth’’ with “decline’’).

\paragraph{Temporal Semantic Volatility (TSV).}
Quantifies the instability of semantic representations across time, reflecting narrative drift between consecutive periods:
\begin{equation}
\text{TSV} = \frac{1}{N - 1} \sum_{i=1}^{N-1} \big\| \phi(x_{i+1}) - \phi(x_i) \big\|_2,
\end{equation}
where $\phi(x)$ denotes a sentence embedding.

\paragraph{NLI-based Logical Consistency Score (NLICS).}
Evaluates whether predictions are logically coherent with the news narrative using natural language inference:
\[
H_i = \begin{cases}
\text{``Stock price will increase''}, & f_\theta(x_i) > 0, \\
\text{``Stock price will decrease''}, & \text{otherwise}.
\end{cases}
\]
\begin{equation}
\text{NLICS} = \frac{1}{N} \sum_{i=1}^{N} \operatorname{EntailmentScore}(x_i, H_i).
\end{equation}
This captures narrative plausibility from an entailment perspective.

\subsection{Robustness Profile}
The regime-level robustness profile is defined as
\begin{equation}
\mathcal{M}_{r_k}(f_\theta) = \Big[ \mathcal{L}_{\text{MSE}}, \text{FCAS}, \text{PCS}, \text{TSV}, \text{NLICS} \Big].
\end{equation}
Comparing $\mathcal{M}_{r_k}$ across regimes highlights when a model is stable, when it is fragile, and which sources of drift (causal, perturbation, semantic, logical) drive prediction errors.

\section{Experimental Setup}
We frame the task as a regression problem: given a financial news article and optional metadata (e.g., TF-IDF features, embeddings, or sector identifiers), the model predicts the next-day stock return, denoted as \texttt{Movement\%}. Formally, the model learns a function $f_\theta: x_i \rightarrow \hat{y}_i$, where predictive performance is evaluated via mean squared error (MSE):

\begin{equation}
\text{MSE} = \frac{1}{N} \sum_{i=1}^N (\hat{y}_i - y_i)^2
\end{equation}

Lower values indicate higher predictive accuracy and robustness.

\subsection{Data Preprocessing}
To capture recent economic volatility, we construct a temporally segmented dataset spanning \textbf{2018 to 2023}, covering key macro-financial shifts including the COVID-19 pandemic and the rate hike cycle. We select 110 S\&P 500 companies, equally sampled across 11 GICS sectors, and align daily stock returns with timestamped news from the FNSPID dataset~\cite{dong2024fnspid}.

\begin{itemize}
    \item \textbf{News–Price Alignment:} Articles are timestamp-matched to next-day returns.
    \item \textbf{Text Processing:} We apply TF-IDF vectorisation (max features = 2000) and extract sentence embeddings using \\ 
    \texttt{all-MiniLM-L6-v2}.
    \item \textbf{Regime Assignment:} Data are segmented into four regimes: pre-COVID, COVID, post-COVID, and rate-hike.
    \item \textbf{Chronological Splits:} For each stock, data are divided into training (60\%), validation (20\%), and test (20\%) sets by timestamp.
\end{itemize}
The regime windows are as follows:

\begin{table}[t]
    \centering
        \caption{Temporal regime windows used for regime-aware evaluation.}
    \renewcommand{\arraystretch}{2.0}
    \setlength{\tabcolsep}{6pt}
    \begin{tabular}{lcc}
        \toprule
        \textbf{Regime} & \textbf{Start Date} & \textbf{End Date} \\
        \midrule
        Pre-COVID  & 2019-11-01 & 2019-12-31 \\
        COVID      & 2020-01-01 & 2020-03-23 \\
        Post-COVID & 2020-05-01 & 2020-07-01 \\
        Rate-Hike  & 2022-02-15 & 2022-06-15 \\
        \bottomrule
    \end{tabular}

    \label{tab:regime_dates}
\end{table}

\subsection{Sector Composition}
To ensure balanced evaluation across economic and industry narratives, we selected 110 S\&P 500 companies equally sampled from 11 GICS sectors. This ensures diversity across technology, healthcare, finance, and consumer verticals.

% \begin{table}[h]
% \centering
% \small
% \renewcommand{\arraystretch}{1.15}
% \setlength{\tabcolsep}{3pt}
% \begin{tabular}{lp{10.5cm}} 
% \toprule
% \textbf{Sector} & \textbf{Selected Companies} \\
% \midrule
% Information Technology & AAPL, ADBE, AMD, ACN, ADSK, CSCO, ADI, AMAT, ANSS, HPE \\
% Health Care & ABT, ABBV, BMY, BIIB, JNJ, LLY, AMGN, CVS, CI, HUM \\
% Financials & AXP, BAC, JPM, GS, MS, BLK, AIG, ALL, MET, C \\
% Consumer Discretionary & AMZN, TSLA, HD, NKE, LOW, TJX, CMG, BKNG, ULTA, ROST \\
% Communication Services & GOOGL, GOOG, T, DIS, CMCSA, CHTR, NFLX, FOXA, FOX, TMUS \\
% Industrials & BA, CAT, GE, MMM, GD, DE, ETN, EMR, DOV, JBHT \\
% Consumer Staples & KO, PEP, PM, CL, GIS, CPB, KR, SYY, CAG, CLX \\
% Energy & XOM, CVX, PSX, COP, SLB, HAL, MPC, VLO, OKE, APA \\
% Utilities & DUK, AEP, NEE, SO, EXC, ED, XEL, EIX, PPL, AES \\
% Real Estate & ARE, AMT, AVB, CCI, DLR, EQIX, O, EXR, PSA, BXP \\
% Materials & APD, ALB, DD, NUE, LIN, DOW, FMC, AVY, NEM, MLM \\
% \bottomrule
% \end{tabular}
% \caption{110 companies selected from the FNSPID dataset~\cite{dong2024fnspid}, categorised by GICS sector.}
% \label{tab:fns_pid_companies}
% \end{table}

\begin{table*}[t]
\centering
\small
\renewcommand{\arraystretch}{1.15}
\setlength{\tabcolsep}{3pt}
\begin{tabular}{l@{\hspace{14pt}}p{13.1cm}} 
\toprule
\textbf{Sector} & \textbf{Selected Companies} \\
\midrule
Information Technology & AAPL, ADBE, AMD, ACN, ADSK, CSCO, ADI, AMAT, ANSS, HPE \\
Health Care & ABT, ABBV, BMY, BIIB, JNJ, LLY, AMGN, CVS, CI, HUM \\
Financials & AXP, BAC, JPM, GS, MS, BLK, AIG, ALL, MET, C \\
Consumer Discretionary & AMZN, TSLA, HD, NKE, LOW, TJX, CMG, BKNG, ULTA, ROST \\
Communication Services & GOOGL, GOOG, T, DIS, CMCSA, CHTR, NFLX, FOXA, FOX, TMUS \\
Industrials & BA, CAT, GE, MMM, GD, DE, ETN, EMR, DOV, JBHT \\
Consumer Staples & KO, PEP, PM, CL, GIS, CPB, KR, SYY, CAG, CLX \\
Energy & XOM, CVX, PSX, COP, SLB, HAL, MPC, VLO, OKE, APA \\
Utilities & DUK, AEP, NEE, SO, EXC, ED, XEL, EIX, PPL, AES \\
Real Estate & ARE, AMT, AVB, CCI, DLR, EQIX, O, EXR, PSA, BXP \\
Materials & APD, ALB, DD, NUE, LIN, DOW, FMC, AVY, NEM, MLM \\
\bottomrule
\end{tabular}
\caption{110 companies selected from the FNSPID dataset~\cite{dong2024fnspid}, categorised by GICS sector.}
\label{tab:fns_pid_companies}
\end{table*}

\subsection{Experimental Pipeline}
The complete regime-aware evaluation procedure is summarised in Algorithm~\ref{alg:pipeline}.

\begin{algorithm}[ht]
\caption{Regime-Aware Evaluation Pipeline}
\label{alg:pipeline}
\begin{algorithmic}[1]
\Require Dataset $\mathcal{D}$, regime partitions $\mathcal{R}$, models $\{f_\theta\}$, metrics $\{\mathcal{M}\}$
\Ensure Robustness profiles $\mathcal{M}_{r}(f_\theta)$ for all $r \in \mathcal{R}$

\State Preprocess financial news: extract TF-IDF, MiniLM embeddings, align with returns
\For{each regime $r \in \mathcal{R}$}
    \State Define regime-specific data $\mathcal{D}_r \subset \mathcal{D}$
    \State Chronologically split $\mathcal{D}_r$ into train/val/test sets
    \For{each model $f_\theta$}
        \State Train $f_\theta$ on $\mathcal{D}_r^{\text{train}}$ using MSE loss
        \State Evaluate $f_\theta$ on $\mathcal{D}_r^{\text{test}}$
        \State Compute metrics: MSE, FCAS, PCS, TSV, NLICS
        \State Save robustness profile $\mathcal{M}_{r}(f_\theta)$
    \EndFor
\EndFor
\end{algorithmic}
\end{algorithm}

\subsection{System Setup and GPT-4 Evaluation Details}
\paragraph{LLM-Based Evaluation (NLICS).}
To compute the News-Logic Inference Coherence Score (NLICS), we use GPT-4 to evaluate the logical alignment between financial news and model predictions. Each prompt follows the format:

\begin{quote}
\texttt{News: "[News excerpt]"\\
Prediction: "Stock will rise"\\
Question: Is the prediction logically supported by the news?\\
Answer: [Yes/No/Uncertain]}
\end{quote}

Responses are mapped to scores: “Yes” with >80\% confidence is scored as 1.0, “Uncertain” as 0.5, and “No” as 0. GPT-4 (April 2024 version) was accessed via the OpenAI API.

\paragraph{System Setup.}
Experiments were conducted on a workstation with a single NVIDIA A6000 GPU (CUDA 12.6) and 256GB RAM. Sentence embeddings were extracted using \texttt{all-MiniLM-L6-v2} via HuggingFace Transformers.

\paragraph{Data Sources.}
\begin{itemize}
    \item \textbf{News:} Financial articles from the FNSPID dataset~\cite{dong2024fnspid}.
    \item \textbf{Returns:} Daily stock prices from Yahoo Finance (2018–2023).
    \item \textbf{Models:} Implemented in PyTorch (LSTM, Transformer variants).
    \item \textbf{NLI Evaluation:} GPT-4 and BART-NLI (HuggingFace Transformers).
\end{itemize}
\subsection{Baseline Models}
We compare the following model classes:
\begin{itemize}
    \item \textbf{LSTM-based models}~\cite{hochreiter1997long}: Trained on TF-IDF representations.
    \item \textbf{Text-only Transformers}~\cite{sanh2019distilbert}: DistilBERT fine-tuned on raw news content.
    \item \textbf{Feature-augmented Transformers}~\cite{vaswani2017attention}: Trained on concatenated TF-IDF vectors and MiniLM embeddings.
\end{itemize}
All models use a learning rate of 0.001, batch size of 64, hidden size of 256, and dropout of 0.2. Training is conducted using the Adam optimiser.

\section{Results}
We evaluate model robustness under macroeconomic regime shifts using our proposed metrics. Results cover LSTM, Transformer, and feature enhanced models, highlighting performance under semantic drift and the effect of causal and logical alignment. We also include ablation studies to assess the contribution of each component.

\subsection{Performance Across Economic Regimes}
We evaluate model robustness under four distinct macro-financial regimes using Mean Squared Error (MSE). Results in Table~\ref{tab:results} highlight how performance varies across periods of stability, crisis, and recovery.

\begin{table}[ht]
\centering
\caption{Mean squared error (MSE) of models across economic regimes. Lower is better. Bold indicates the best-performing model per regime.}
\renewcommand{\arraystretch}{1.1}
\setlength{\tabcolsep}{6pt}
\resizebox{\linewidth}{!}{%
\begin{tabularx}{\linewidth}{lXXX}
\toprule
\textbf{Regime} & \textbf{LSTM (MSE ↓)} & \textbf{Text Transformer (MSE ↓)} & \textbf{Feature Transformer (MSE ↓)} \\
\midrule
Pre-COVID    & 3.08 & \textbf{2.80} & 3.19 \\
COVID        & \textbf{3.74} & 40.95 & 32.02 \\
Post-COVID   & \textbf{3.48} & 4.44 & 3.76 \\
Rate-Hike    & 6.47 & \textbf{5.09} & 7.01 \\
\bottomrule
\end{tabularx}
}
\label{tab:results}
\end{table}

\noindent\textbf{Summary.}
LSTM maintains consistent performance with low regime-induced variance (std.~$\approx$~1.34). In contrast, the Text Transformer exhibits severe degradation during COVID, indicating high sensitivity to semantic drift. The Feature Transformer shows improved post-COVID stability, highlighting the benefit of incorporating structured financial indicators.

\subsection{Semantic Drift and Vocabulary Shift} 
\label{semanticshift}
\begin{figure}[ht] 
    \centering 
    \includegraphics[width=0.8\linewidth]{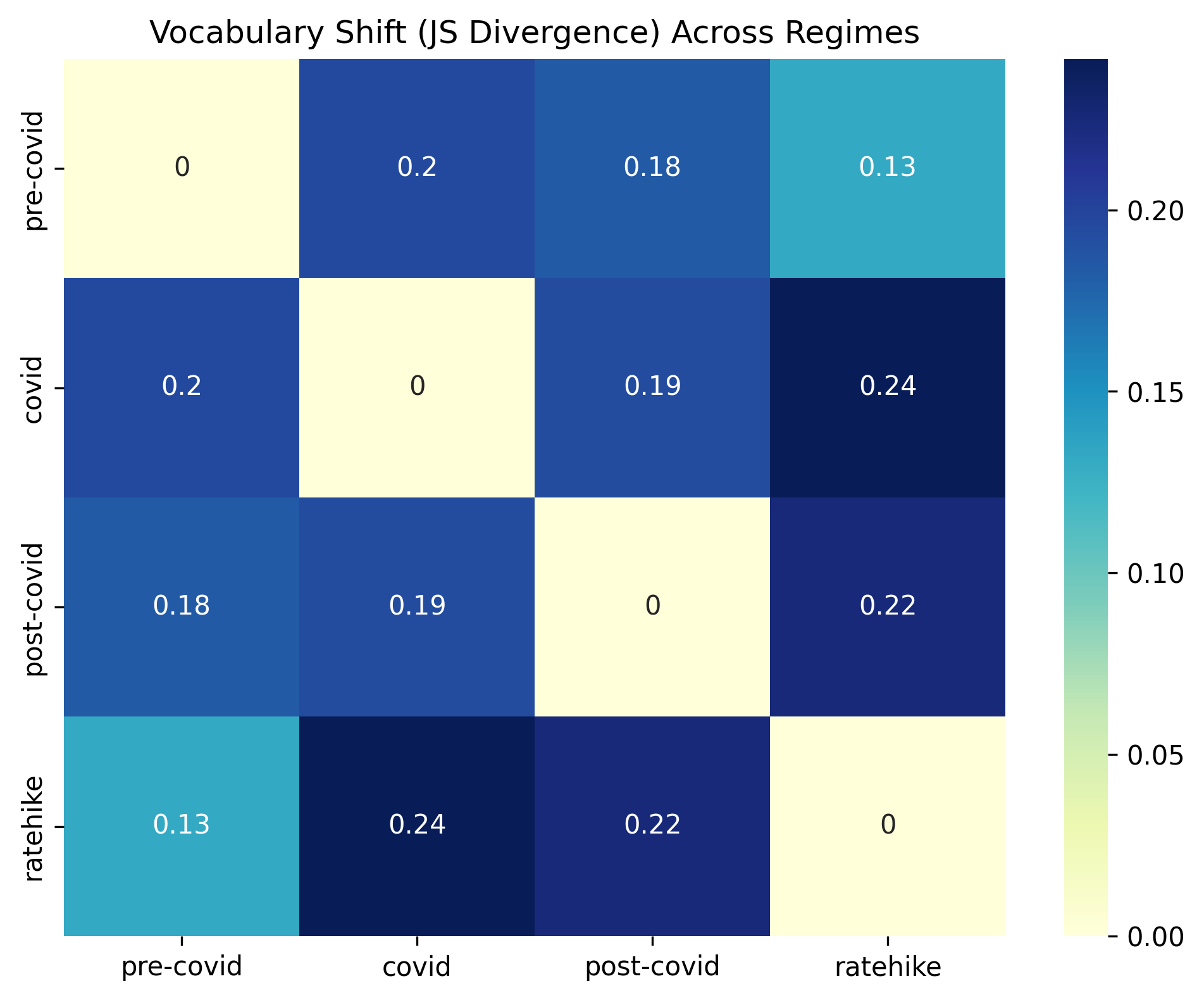} 
    \caption{Jensen-Shannon Divergence Between Regime-Specific TF-IDF Distributions}
    \Description{Heatmap where both rows and columns correspond to different macroeconomic regimes. 
    Each cell encodes the Jensen–Shannon divergence between the TF–IDF word distributions of the two regimes. 
    Darker colours indicate higher divergence, meaning the vocabularies and term frequencies differ more strongly 
    between those regimes. The diagonal cells are near zero, showing minimal divergence within the same regime.}
    \label{fig:js_divergence} 
\end{figure}

To measure linguistic change, we compute Jensen-Shannon (JS) divergence between TF-IDF distributions of news across regimes. Figure~\ref{fig:js_divergence} shows that the greatest shift occurs between \textbf{COVID and rate-hike} (JS = 0.24), aligning with peak model instability. Pairs such as \textit{pre-COVID vs. COVID} (0.20) and \textit{post-COVID vs. rate-hike} (0.22) also exhibit substantial divergence, reflecting macro-financial discontinuities.

\subsection{Sector Transferability}
We assess cross-domain robustness by training on financial-sector news and testing on healthcare. Table~\ref{tab:sector_transfer} reports MSE across model types. Feature-based models yield higher test error but better generalisability, relying less on domain-specific terms.

\begin{table}[ht]
\centering
\caption{Cross-sector MSE (trained on financial, tested on healthcare). Feature-based models generalise better despite higher error.}
\begin{tabular}{lcc}
\toprule
\textbf{Model} & \textbf{Text-Only MSE} & \textbf{Feature-Based MSE} \\
\midrule
MiniLM & \textbf{0.198} & 0.508 \\
MPNet  & \textbf{0.201} & 0.486 \\
Raw Text & \textbf{0.201} & 0.469 \\
\bottomrule
\end{tabular}
\label{tab:sector_transfer}
\end{table}

\subsection{Visualising Regime-Aware Representations.}
\begin{figure}[ht] 
    \centering 
    \includegraphics[width=1.05\linewidth]{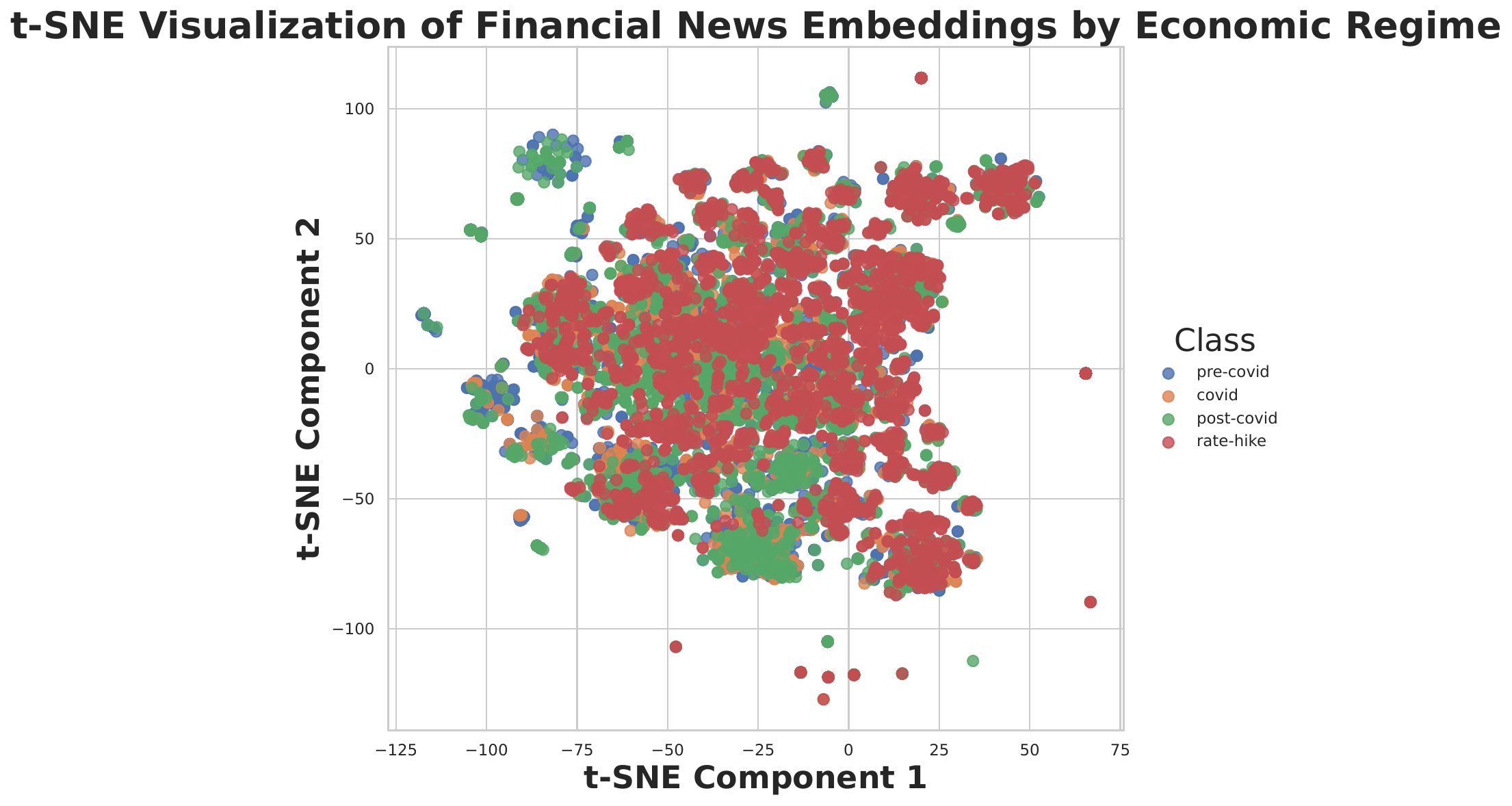} 
    \caption{t-SNE visualisation of financial news embeddings by economic regime.}
    \Description{Scatter plot showing a two-dimensional t-SNE projection of news embeddings. 
    Points are coloured by macroeconomic regime, forming distinct clusters that indicate separability 
    of embedding space with respect to regimes.}
    \label{fig:tsne_regime} 
\end{figure}

\begin{figure}[ht] 
    \centering 
    \includegraphics[width=1.05\linewidth]{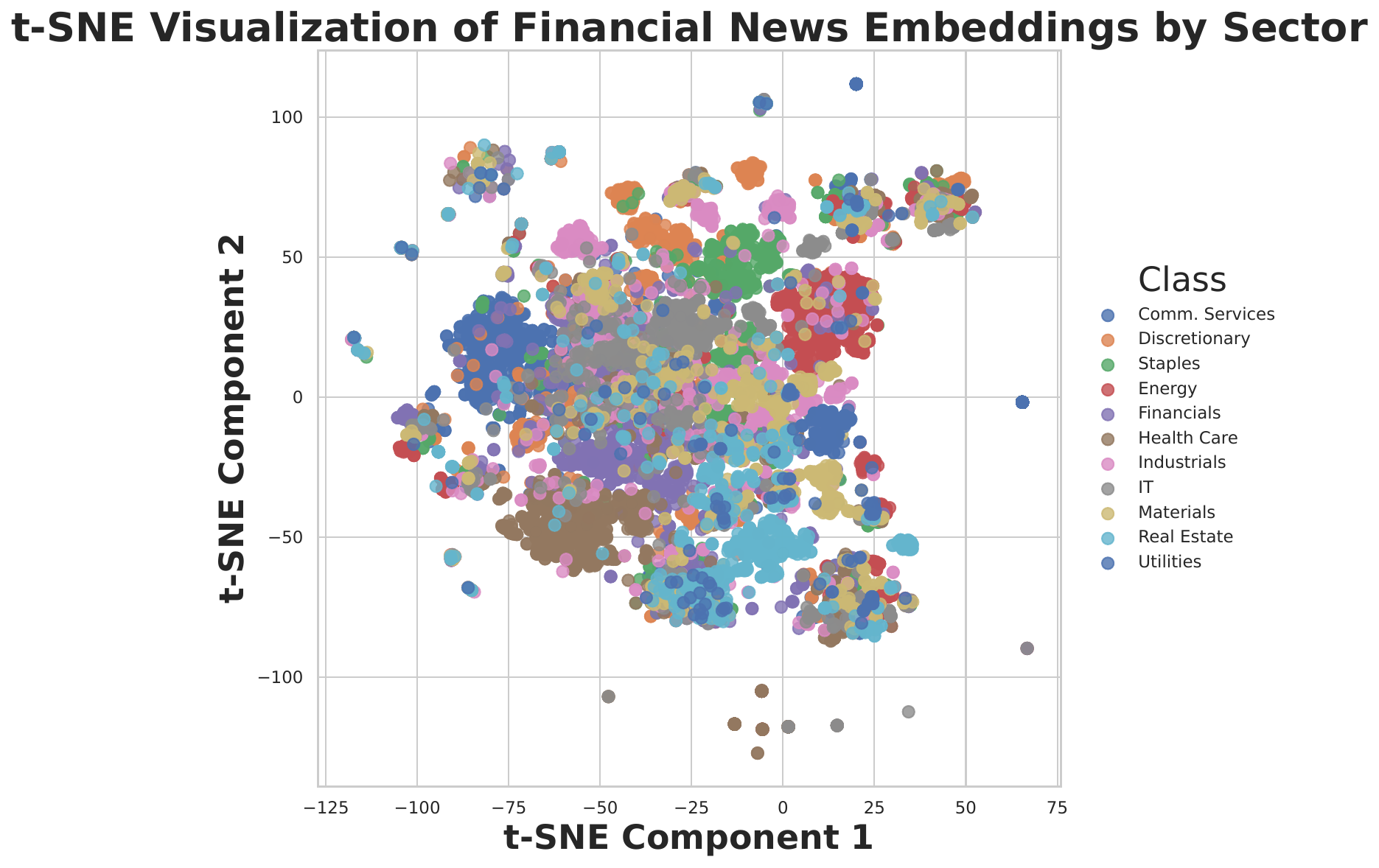} 
    \caption{t-SNE visualisation of financial news embeddings by industry sector.}
    \Description{Scatter plot showing a two-dimensional t-SNE projection of news embeddings. 
    Points are coloured by industry sector. Clusters correspond to different sectors, 
    highlighting how embeddings capture sector-specific patterns.}
    \label{fig:tsne_sector} 
\end{figure}

We examine whether learned model representations capture temporal or sectoral distinctions by projecting sentence-level embeddings using t-SNE. Figure~\ref{fig:tsne_regime} shows that economic regimes exhibit some separability, particularly for \textit{pre-COVID} and \textit{rate-hike} periods, suggesting partial regime-awareness in the embedding space. In contrast, Figure~\ref{fig:tsne_sector} reveals sharper, more coherent clustering by industry sector, especially for \textit{Information Technology}, \textit{Health Care}, and \textit{Financials}. This indicates that model representations encode sectoral narratives more robustly than macroeconomic temporal structure.

\paragraph{Key Findings.}
\begin{itemize}
    \item \textbf{Semantic Shift Across Regimes:} Jensen-Shannon divergence peaks at \textbf{0.24} (pre-COVID vs. COVID) and \textbf{0.27} (rate-hike vs. COVID), aligning with increased model error and confirming the impact of distributional shift.
    
    \item \textbf{LSTM vs. Transformer Dynamics:} LSTMs exhibit more consistent performance across regimes. Transformers, while more volatile, show stronger \textbf{logical alignment} as captured by NLICS, indicating their capacity to model complex language patterns.
    
    \item \textbf{Sector Generalisation via Feature-Augmented Models:} Feature-based Transformers achieve improved sector-level robustness and lower semantic volatility (TSV), making them more resilient under structural variation.
\end{itemize}

\subsection{Controls for Situational vs. Linguistic Drift}
While Section~\ref{semanticshift} shows that regime shifts correlate with vocabulary divergence and volatility, it remains unclear whether this instability arises from (i) fundamentally different economic situations (e.g., crisis vs.\ recovery) or (ii) changes in how similar events are described across regimes. To disentangle these effects, we design two control experiments.

\paragraph{Within-Regime Event Controls.}
We first examine comparable financial events within the same regime, such as quarterly earnings announcements. For each event class, we compute PCS and TSV scores across firms. Consistently low volatility in this setting indicates that instability is primarily situational rather than linguistic.

\paragraph{Cross-Regime Matched Events.}
We then compare the same type of events across regimes (e.g., earnings announcements pre-COVID vs.\ during COVID). Significant increases in TSV and PCS suggest that models face true narrative drift in how events are framed, even when the economic signal is similar.

\paragraph{Findings.}
Table~\ref{tab:controls} reports results for earnings-related events. Within-regime comparisons yield relatively low volatility (TSV $\approx 0.9$), whereas cross-regime matched events exhibit substantially higher volatility (TSV $\approx 1.8$). This pattern confirms that both situational differences and linguistic drift contribute to robustness degradation, with narrative re-framing across regimes amplifying fragility.

\begin{table}[ht]
\centering
\caption{Perturbation sensitivity (PCS) and semantic volatility (TSV) for matched earnings events. Lower values indicate higher stability.}
\label{tab:controls}
\begin{tabular}{lcc}
\toprule
Event Type & PCS $\downarrow$ & TSV $\downarrow$ \\
\midrule
Within-Regime (pre-COVID) & 1.21 & 0.92 \\
Within-Regime (COVID)     & 1.35 & 0.87 \\
Cross-Regime (pre-COVID vs.\ COVID) & 2.04 & 1.82 \\
Cross-Regime (post-COVID vs.\ rate-hike) & 1.89 & 1.74 \\
\bottomrule
\end{tabular}
\end{table}

\subsection{Case Study: Metric Performance Across Regimes}
To better understand model robustness at the stock level, we apply our four evaluation metrics to two representative firms: JPMorgan Chase (JPM) and Apple (AAPL), across all four economic regimes. We use GPT-4 as an auxiliary evaluator to estimate entailment between predicted movement and input news. Each prompt follows the structure:

\texttt{``Given the financial news: [NEWS], and the predicted outcome: [UP/DOWN], is the prediction logically supported by the news?''}

GPT-4's responses are mapped to numeric scores using calibrated entailment thresholds (details in Appendix~A.3). Table~\ref{tab:case_metrics} presents the FCAS, PCS, TSV, and NLICS scores for both stocks.

\begin{table}[ht]
\centering
\caption{Diagnostic metric values for JPM and AAPL across economic regimes.}
\small
\resizebox{\linewidth}{!}{
\begin{tabular}{c l c c c c}
\toprule
\textbf{Stock} & \textbf{Regime} & \textbf{FCAS} & \textbf{PCS} & \textbf{TSV} & \textbf{NLICS} \\
\midrule
\multirow{4}{*}{JPM} & Pre-COVID   & 1.118 & 2.939 & 1.703 & 0.60 \\
                     & COVID       & -2.096 & -0.824 & 2.10 & 0.45 \\
                     & Post-COVID  & -0.948 & 2.946 & 1.60 & 0.53 \\
                     & Rate-Hike   & 2.137 & 2.899 & 1.85 & 0.55 \\
\midrule
\multirow{4}{*}{AAPL} & Pre-COVID   & 1.146 & 1.722 & 1.15 & 0.66 \\
                      & COVID       & -2.090 & 2.836 & 2.30 & 0.40 \\
                      & Post-COVID  & -0.880 & -0.916 & 1.70 & 0.58 \\
                      & Rate-Hike   & 2.153 & 1.063 & 1.90 & 0.57 \\
\bottomrule
\end{tabular}
}

\label{tab:case_metrics}
\end{table}

\paragraph{Interpretation.}
\begin{itemize}
    \item \textbf{FCAS:} Causal alignment drops during COVID for both stocks, consistent with disrupted macro-financial narratives. Scores rebound in the rate-hike period, suggesting recovery in interpretability.
    \item \textbf{PCS:} JPM shows maximum perturbation sensitivity pre-COVID (2.939), possibly reflecting fragility to regulatory tone. AAPL peaks during COVID (2.836), consistent with heightened uncertainty in the tech sector.
    \item \textbf{TSV:} Semantic drift peaks during COVID (JPM: 2.10, AAPL: 2.30), validating the volatility captured by our distributional analysis.
    \item \textbf{NLICS:} Logical coherence is lowest during COVID (JPM: 0.45, AAPL: 0.40), confirming GPT-4’s identification of alignment breakdown between input and prediction.
\end{itemize}

\section{Ablation Study: Impact of Causal and Semantic Components}
We conduct an ablation study to evaluate the role of each metric and model component in contributing to robustness under regime shifts. Specifically, we measure the impact of removing each diagnostic metric individually, and assess the effects of feature augmentation and entailment model choice.

\subsection{Metric-Specific Ablations}
Table~\ref{tab:ablation_metrics} reports results for a Transformer model evaluated during the rate hike period. Each row corresponds to the removal of a specific metric from the evaluation pipeline. Metrics marked as \textit{N/A} indicate that the corresponding score was not computed, as the component was excluded from that configuration. Removing FCAS or NLICS produces the largest reductions in interpretability.

\begin{table}[ht]
\centering
\caption{
Ablation of diagnostic metrics. Each variant excludes one metric from the evaluation. \textit{N/A} indicates the metric was omitted from both computation and scoring in that configuration.
}
\renewcommand{\arraystretch}{1.1}
\setlength{\tabcolsep}{6pt}
\resizebox{\linewidth}{!}{%
\begin{tabularx}{\linewidth}{lXXXX}
\toprule
\textbf{Model Variant} & \textbf{FCAS} & \textbf{PCS} & \textbf{TSV} & \textbf{NLICS} \\
\midrule
Full Evaluation        & 0.62 & 2.91 & 1.78 & 0.56 \\
No FCAS                & \textit{N/A} & 3.07 & 1.85 & 0.49 \\
No PCS                 & 0.60 & \textit{N/A} & 1.72 & 0.53 \\
No TSV                 & 0.58 & 2.90 & \textit{N/A} & 0.55 \\
No NLICS               & 0.61 & 2.94 & 1.76 & \textit{N/A} \\
\bottomrule
\end{tabularx}
}
\label{tab:ablation_metrics}
\end{table}

\subsection{Effect of Feature Augmentation}
We compare models trained on raw text alone with those enhanced using TF-IDF and MiniLM features. Table~\ref{tab:ablation_features} shows that feature-enhanced models achieve lower semantic volatility and higher logical consistency, indicating improved robustness under drift.

\begin{table}[ht]
\centering
\caption{Effect of feature augmentation on semantic drift, logical consistency, and generalisation.}
\renewcommand{\arraystretch}{1.1}
\setlength{\tabcolsep}{6pt}
\resizebox{\linewidth}{!}{%
\begin{tabularx}{\linewidth}{lXXX}
\toprule
\textbf{Model Type} & \textbf{TSV} & \textbf{NLICS} & \textbf{Cross-Sector MSE} \\
\midrule
Text Only           & 2.07 & 0.49 & 0.501 \\
Feature Enhanced    & \textbf{1.76} & \textbf{0.56} & \textbf{0.469} \\
\bottomrule
\end{tabularx}
}
\label{tab:ablation_features}
\end{table}

\subsection{Comparison of Entailment Models}
We evaluate NLICS using both BART-NLI and GPT-4. As shown in Table~\ref{tab:ablation_llm}, GPT-4 achieves higher agreement with human judgment and better captures nuanced entailment, though it is computationally more expensive.

\begin{table}[ht]
\centering
\caption{Comparison of entailment models for computing NLICS. GPT-4 shows better alignment with expert-labeled ground truth.}
\label{tab:ablation_llm}
\renewcommand{\arraystretch}{1.1}
\setlength{\tabcolsep}{6pt}
\resizebox{0.7\linewidth}{!}{%
\begin{tabular}{lrr}
\toprule
\textbf{Entailment Model} & \textbf{NLICS} & \textbf{Human Agreement (\%)} \\
\midrule
BART-NLI & 0.52 & 72.1 \\
GPT-4    & \textbf{0.56} & \textbf{85.6} \\
\bottomrule
\end{tabular}%
}
\end{table}

\section{Discussion}
Our findings highlight the difficulty of maintaining predictive reliability in financial NLP under changing economic conditions. Performance degrades notably during periods of crisis, with models struggling to adapt to evolving narratives and causal structures.
The proposed metrics offer complementary insights: FCAS and NLICS capture alignment with causal and logical content, while PCS and TSV reflect sensitivity to perturbation and semantic drift. Together, they provide a more complete view of robustness than accuracy alone. LSTM models show greater stability across regimes, while Transformers offer higher expressiveness but are more affected by drift. Feature augmentation improves generalisation and interpretability, suggesting a trade-off between flexibility and robustness. Our GPT-4 case study shows that large language models can support post-hoc auditing, though their integration introduces new challenges in cost and transparency. Overall, the results support the need for diagnostic evaluation tools in building reliable, regime-aware financial prediction systems.

\section{Conclusion}
We propose a regime-aware evaluation framework for financial natural language processing that quantifies model robustness under temporal, causal, and semantic shift. The framework introduces four diagnostic metrics: Financial Causal Attribution Score (FCAS), Patent Cliff Sensitivity (PCS), Temporal Semantic Volatility (TSV), and NLI-based Logical Consistency Score (NLICS), each designed to assess a distinct aspect of predictive reliability.Our empirical results show that LSTM models exhibit greater stability across economic regimes, while Transformer models, though more expressive, are more vulnerable to shifts in financial narratives. Feature-enhanced Transformers demonstrate improved generalisation and reduced volatility, particularly in post-crisis periods. These findings highlight the importance of regime-aware auditing for financial prediction systems. Our framework supports early identification of failure modes and promotes the development of adaptive, interpretable, and robust AI systems in dynamic market environments.Future directions include extending the framework to multimodal inputs such as earnings calls and investor briefings, incorporating real-time feedback for continual adaptation, and applying the methodology to reinforcement learning scenarios in financial decision-making.

\printbibliography

@String{Computing = "Computing" }

@String{Computer = "{IEEE} Computer" }

@BOOK{test,
   author = "Donald E. Knuth",
   title = "Seminumerical Algorithms",
   volume = 2,
   series = "The Art of Computer Programming",
   publisher = "Addison-Wesley",
   address = "Reading, MA",
   edition = "2nd",
   month = "10~" # jan,
   year = "1981",
}

@ArtifactSoftware{R,
    title = {R: A Language and Environment for Statistical Computing},
    author = {{R Core Team}},
    organization = {R Foundation for Statistical Computing},
    address = {Vienna, Austria},
    year = {2019},
    url = {https://www.R-project.org/},
}

@inproceedings{ding2015deep,
  title={Deep learning for event-driven stock prediction},
  author={Ding, Xiao and Zhang, Yue and Liu, Ting and Duan, Junwen},
  booktitle={Twenty-fourth international joint conference on artificial intelligence},
  year={2015}
}

@article{du2024financial,
  title={Financial sentiment analysis: Techniques and applications},
  author={Du, Kelvin and Xing, Frank and Mao, Rui and Cambria, Erik},
  journal={ACM Computing Surveys},
  volume={56},
  number={9},
  pages={1--42},
  year={2024},
  publisher={ACM New York, NY}
}

@inproceedings{dong2024fnspid,
  title={Fnspid: A comprehensive financial news dataset in time series},
  author={Dong, Zihan and Fan, Xinyu and Peng, Zhiyuan},
  booktitle={Proceedings of the 30th ACM SIGKDD Conference on Knowledge Discovery and Data Mining},
  pages={4918--4927},
  year={2024}
}

@article{araci2019finbert,
  title={FinBERT: Financial Sentiment Analysis with Pre-trained Language Models},
  author={Araci, D},
  journal={arXiv preprint arXiv:1908.10063},
  year={2019}
}

@article{wu2023bloomberggpt,
  title={Bloomberggpt: A large language model for finance},
  author={Wu, Shijie and Irsoy, Ozan and Lu, Steven and Dabravolski, Vadim and Dredze, Mark and Gehrmann, Sebastian and Kambadur, Prabhanjan and Rosenberg, David and Mann, Gideon},
  journal={arXiv preprint arXiv:2303.17564},
  year={2023}
}

@article{sinha2022sentfin,
  title={SEntFiN 1.0: Entity-aware sentiment analysis for financial news},
  author={Sinha, Ankur and Kedas, Satishwar and Kumar, Rishu and Malo, Pekka},
  journal={Journal of the Association for Information Science and Technology},
  volume={73},
  number={9},
  pages={1314--1335},
  year={2022},
  publisher={Wiley Online Library}
}

@article{gururangan2020don,
  title={Don't stop pretraining: Adapt language models to domains and tasks},
  author={Gururangan, Suchin and Marasovi{\'c}, Ana and Swayamdipta, Swabha and Lo, Kyle and Beltagy, Iz and Downey, Doug and Smith, Noah A},
  journal={arXiv preprint arXiv:2004.10964},
  year={2020}
}

@article{ghosh2024logical,
  title={Logical Consistency of Large Language Models in Fact-checking},
  author={Ghosh, Bishwamittra and Hasan, Sarah and Arafat, Naheed Anjum and Khan, Arijit},
  journal={arXiv preprint arXiv:2412.16100},
  year={2024}
}

@article{wu2020outbreak,
  title={The outbreak of COVID-19: An overview},
  author={Wu, Yi-Chi and Chen, Ching-Sung and Chan, Yu-Jiun},
  journal={Journal of the Chinese medical association},
  volume={83},
  number={3},
  pages={217--220},
  year={2020},
  publisher={LWW}
}

@article{vaswani2017attention,
  title={Attention is all you need},
  author={Vaswani, A},
  journal={Advances in Neural Information Processing Systems},
  year={2017}
}

@article{hochreiter1997long,
  title={Long Short-term Memory},
  author={Hochreiter, S},
  journal={Neural Computation MIT-Press},
  year={1997}
}

@article{sanh2019distilbert,
  title={DistilBERT, a distilled version of BERT: smaller, faster, cheaper and lighter},
  author={Sanh, V},
  journal={arXiv preprint arXiv:1910.01108},
  year={2019}
}

@inproceedings{priya2025advanced,
  title={Advanced financial sentiment analysis using FinBERT to explore sentiment dynamics},
  author={Priya, S Baghavathi and Kumar, Madhav and JD, Nitheesh Prakash and others},
  booktitle={2025 3rd International Conference on Intelligent Data Communication Technologies and Internet of Things (IDCIoT)},
  pages={889--897},
  year={2025},
  organization={IEEE}
}

@inproceedings{mahendran2025comparative,
  title={Comparative Advances in Financial Sentiment Analysis: A Review of BERT, FinBert, and Large Language Models},
  author={Mahendran, Manish Barath and Gokul, Aswin Kumar and Lakshmi, Poornima and Pavithra, S},
  booktitle={2025 3rd International Conference on Intelligent Data Communication Technologies and Internet of Things (IDCIoT)},
  pages={39--45},
  year={2025},
  organization={IEEE}
}

@article{kang2025comparative,
  title={Comparative investigation of gpt and finbert’s sentiment analysis performance in news across different sectors},
  author={Kang, Ji-Won and Choi, Sun-Yong},
  journal={Electronics},
  volume={14},
  number={6},
  pages={1090},
  year={2025},
  publisher={MDPI}
}

@inproceedings{jun2024predicting,
  title={Predicting stock prices with finbert-lstm: Integrating news sentiment analysis},
  author={jun Gu, Wen and hao Zhong, Yi and zun Li, Shi and song Wei, Chang and ting Dong, Li and yue Wang, Zhuo and Yan, Chao},
  booktitle={Proceedings of the 2024 8th International Conference on Cloud and Big Data Computing},
  pages={67--72},
  year={2024}
}

@article{thomas2024enhancing,
  title={Enhancing tinybert for financial sentiment analysis using gpt-augmented finbert distillation},
  author={Thomas, Graison Jos},
  journal={arXiv preprint arXiv:2409.18999},
  year={2024}
}

@article{jin2025early,
  title={An early prediction model on systemic risk under global risk: Using FinBERT and temporal fusion transformer to multimodal data fusion framework},
  author={Jin, Xiao and Lin, Shu-Ling},
  journal={The North American Journal of Economics and Finance},
  volume={76},
  pages={102361},
  year={2025},
  publisher={Elsevier}
}

@inproceedings{peng2021domain,
  title={Is domain adaptation worth your investment? comparing BERT and FinBERT on financial tasks},
  author={Peng, Bo and Chersoni, Emmanuele and Hsu, Yu-Yin and Huang, Chu-Ren and others},
  year={2021},
  organization={Association for Computational Linguistics (ACL)}
}

@inproceedings{correa2022neural,
  title={Neural text classification for digital transformation in the financial regulatory domain},
  author={Correa, Nelson and Correa, Antonio},
  booktitle={2022 IEEE ANDESCON},
  pages={1--6},
  year={2022},
  organization={IEEE}
}

@article{sharkey2024bert,
  title={BERT vs GPT for financial engineering},
  author={Sharkey, Edward and Treleaven, Philip},
  journal={arXiv preprint arXiv:2405.12990},
  year={2024}
}

@article{shobayo2024innovative,
  title={Innovative sentiment analysis and prediction of stock price using FinBERT, GPT-4 and logistic regression: A data-driven approach},
  author={Shobayo, Olamilekan and Adeyemi-Longe, Sidikat and Popoola, Olusogo and Ogunleye, Bayode},
  journal={Big Data and Cognitive Computing},
  volume={8},
  number={11},
  pages={143},
  year={2024},
  publisher={MDPI}
}

@inproceedings{gossi2023finbert,
  title={Finbert-fomc: Fine-tuned finbert model with sentiment focus method for enhancing sentiment analysis of fomc minutes},
  author={G{\"o}ssi, Sandro and Chen, Ziwei and Kim, Wonseong and Bermeitinger, Bernhard and Handschuh, Siegfried},
  booktitle={Proceedings of the Fourth ACM International Conference on AI in Finance},
  pages={357--364},
  year={2023}
}

@inproceedings{sun2025ricciflowrec,
  title={RicciFlowRec: A Geometric Root Cause Recommender Using Ricci Curvature on Financial Graphs},
  author={Sun, Zhongtian and Harit, Anoushka},
  booktitle={Proceedings of the Nineteenth ACM Conference on Recommender Systems},
  pages={1284--1289},
  year={2025}
}

@article{sun2025glance,
  title={GLANCE: Graph Logic Attention Network with Cluster Enhancement for Heterophilous Graph Representation Learning},
  author={Sun, Zhongtian and Harit, Anoushka and Cristea, Alexandra and Donnelly, Christl A and Li{\`o}, Pietro},
  journal={arXiv preprint arXiv:2507.18521},
  year={2025}
}

@article{harit2024breaking,
  title={Breaking down financial news impact: A novel ai approach with geometric hypergraphs},
  author={Harit, Anoushka and Sun, Zhongtian and Yu, Jongmin and Moubayed, Noura Al},
  journal={arXiv preprint arXiv:2409.00438},
  year={2024}
}

@article{sun2023money,
  title={Money: Ensemble learning for stock price movement prediction via a convolutional network with adversarial hypergraph model},
  author={Sun, Zhongtian and Harit, Anoushka and Cristea, Alexandra I and Wang, Jingyun and Lio, Pietro},
  journal={AI Open},
  volume={4},
  pages={165--174},
  year={2023},
  publisher={Elsevier}
}

@article{harit2025textfold,
  title={TextFold: A Geometric Hypergraph Framework for Protein Structure Prediction with Scientific Literature Integration},
  author={Harit, Anoushka and Sun, Zhongtian and Al Moubayed, Noura},
  journal={Procedia Computer Science},
  volume={264},
  pages={309--318},
  year={2025},
  publisher={Elsevier}
}

@inproceedings{sun2022contrastive,
  title={Contrastive learning with heterogeneous graph attention networks on short text classification},
  author={Sun, Zhongtian and Harit, Anoushka and Cristea, Alexandra I and Yu, Jialin and Shi, Lei and Al Moubayed, Noura},
  booktitle={2022 International Joint Conference on Neural Networks (IJCNN)},
  pages={1--6},
  year={2022},
  organization={IEEE}
}

\end{document}